\documentclass[amsmath,amssymb,
showpacs,twocolumn, nobibnotes,aps,prb]{revtex4-1}

\usepackage{graphicx}
\usepackage{dcolumn}
\usepackage{bm}
\usepackage{enumerate}
\usepackage{amsmath,amssymb,amsfonts}
\usepackage{verbatim}
\usepackage{color}
\usepackage{mathrsfs}
\usepackage{setspace}
\usepackage{mathtools}
\usepackage{braket}
\usepackage{lipsum}

\newcommand{\be}{\begin{equation}}
\newcommand{\ee}{\end{equation}}
\def\ba{\begin{aligned}}
\def\ea{\end{aligned}}

\newcommand{\bea}{\begin{eqnarray}}
\newcommand{\eea}{\end{eqnarray}}

\begin{document}

\title{Explicit derivation of the chiral and (generic) helical edge states
for the Kane-Mele model:
Closed expressions for the wave function, dispersion relation, and spin rotation}
\author{Fatemeh Rahmati}
\author{Mohsen Amini}
\email{msn.amini@sci.ui.ac.ir}
\author{Morteza Soltani}  
\email{mo.soltani@sci.ui.ac.ir}
\author{Mozhgan  Sadeghizadeh}  

\affiliation{Department of Physics, Faculty of Physics, University
of Isfahan, Isfahan 81746-73441, Iran}

\begin{abstract}

While one of the most important and intriguing features of the topological insulators is the presence of edge states, the closed-form expressions for the edge states of some famous topological models are still lacking. Here, we focus on the Kane-Mele model with and without Rashba spin-orbit coupling as a well-known model to describe a two-dimensional version of the $\mathbb{Z}_2$ topological insulator to study the properties of its edge states analytically. By considering the tight-binding model on a honeycomb lattice with zigzag boundaries and introducing a perturbative approach, we derive explicit expressions for the wave functions, energy dispersion relations, and the spin rotations of the (generic) helical edge states. To this end, we first map the edge states of the ribbon geometry into an effective two-leg ladder model with momentum-dependent energy parameters. Then, we split the Hamiltonian of the system into an unperturbed part and a perturbation. The unperturbed part has a flat-band energy spectrum and can be solved exactly which allows us to consider the remaining part of the Hamiltonian perturbatively. The resulting energy dispersion relation within the first-order perturbation, surprisingly, is in excellent agreement with the numerical spectra over a very wide range of wavenumbers. Our perturbative framework also allows deriving an explicit form for the rotation of the spins of the momentum edge states in the absence of axial spin symmetry due to the Rashba spin-orbit interaction.

\end{abstract}
\maketitle
\section{Introduction~\label{Sec01}}

Since the pioneering description of the Hall coefficient of the integer quantum Hall effect (IQHE) in terms of topological invariants~\cite{Thouless}, the study of topological insulators has always been one of the most fascinating fields of research and innovation in modern condensed matter physics~\cite{Hasan, RMP2011}.
On one side, topological invariants are nonlocal bulk properties of the system which can be fully classified by the symmetries of the Hamiltonian (in the absence of interaction) and allows the distinction of different gapped phases of the system.
On the other side, the presence of such topological invariants is directly related to the emergence of boundary modes (edge states).
In the IQHE, the proper topological invariant is the integral of the $k$-space Berry curvature over the Brillouin zone which is called Chern number and counts the number of gapless chiral edge states of the system~\cite{Thouless, Haldane}.
While the IQHE can occur only when time-reversal symmetry is broken, it is possible to have a topological quantum state in the presence of time-reversal symmetry which is called the quantum spin Hall effect (QSHE). 
The proper topological invariant for characterizing QSHE is the $\mathbb{Z}_2$ invariant that is related to the number of boundary Kramer's pairs localized at the edges in a strip geometry~\cite{KM1, KM2, Bernevig}.

An important model system with fascinating theoretical insights to study two-dimensional QSHE is the well-known Kane-Mele model of graphene with spin-orbit couplings (SOCs)~\cite{KM2}. This model exhibits a pair of helical edge modes with opposite spin-polarization and wave numbers which are connected by time-reversal symmetry.
In the presence of the axial spin symmetry, this model can be considered as two decoupled copies of the Haldane’s model~\cite{Haldane} for electrons with opposite direction of spin.  Therefore, in this regime each helical edge mode can be viewed as a chiral edge mode.
If the axial spin symmetry is broken, by adding the Rashba SOC as a common spin-flipping interaction, spin is no longer a good quantum number. 
In this regime, each momentum eigenstate can generally be a linear combinations of the spin-up and spin-down eigenstates and hence called generic helical mode~\cite{GHS2}. 

It is then surprising that, despite the vast amount of theoretical~\cite{GHS2, AN1, AN2, E1, E2, E3, E4, Amini1, GHS1, GHS3, Oleg} and experimental~\cite{X1, X2, X3, X4, X5, A1} researches on the helical edge states, an explicit closed-form analytic expressions for different properties of such kind of edge states (e.g. wave functions, dispersion relations, and spin rotations) are still lacking.
Analytical investigations so far have been restricted to the use of a combination of analytical and numerical methods. For instance,
the chiral edge states of the Haldane model are studied in Ref.~\cite{AN1} by using the Harper equation to find the wave function transfer relation between two edges in a graphene ribbon. The same method is used to study the edge states of the Kane-Mele model for similar systems~\cite {AN2}.
 
In this paper, we introduce a framework that allows deriving the explicit analytical expressions for the edge states of the Kane-Mele model both in the presence and absence of the Rashba SOC (with and without the axial spin symmetry).
We consider a ribbon geometry of the honeycomb lattice with translational symmetry along the longitudinal direction which allows us to use the corresponding Fourier transform operators and write the Hamiltonian in the momentum space. We then introduce a map between the transformed Hamiltonian and a two-leg ladder structure with momentum-dependent hopping and on-site parameters.  We show that the resulting Hamiltonian for this system can be split into two parts. The first term $\mathcal{H}_0(k_x)$ which could be solved nonperturbatively supports zero-energy edge states with flat bands. The second term $\mathcal{H}_1(k_x)$ can be treated perturbatively to obtain a closed-form expression for the wave functions and energy dispersion relations. Our analysis shows that the first-order perturbation correction is sufficient to achieve an excellent agreement between the resulting solution and their numerical counterparts. We also derive explicitly the rotation of the spin of the momentum eigenstates when the axial spin symmetry breaks in a helical edge state.

The rest of the paper is organized as follows.
In Sec.~\ref{Sec.II}, we describe the topological model employed in this study.
Sec.~\ref{Sec.III} contains a detailed introducing of our perturbative framework to derive the edge states of the system in different symmetry cases for the axial spin symmetry.
This framework requires a map between the edge states of the ribbon geometry and a two-leg ladder model with momentum-dependent hopping and one-site energies which is discussed here. This section also provides the resulting explicit expressions derived using this approach and a comparison with the exact numerical results.
We finally conclude with a summary in Sec.~\ref{Sec.IV}.


\section{Model Hamiltonian}\label{Sec.II}

\begin{figure}[t!]
          \center{\includegraphics[width=1.1\linewidth]{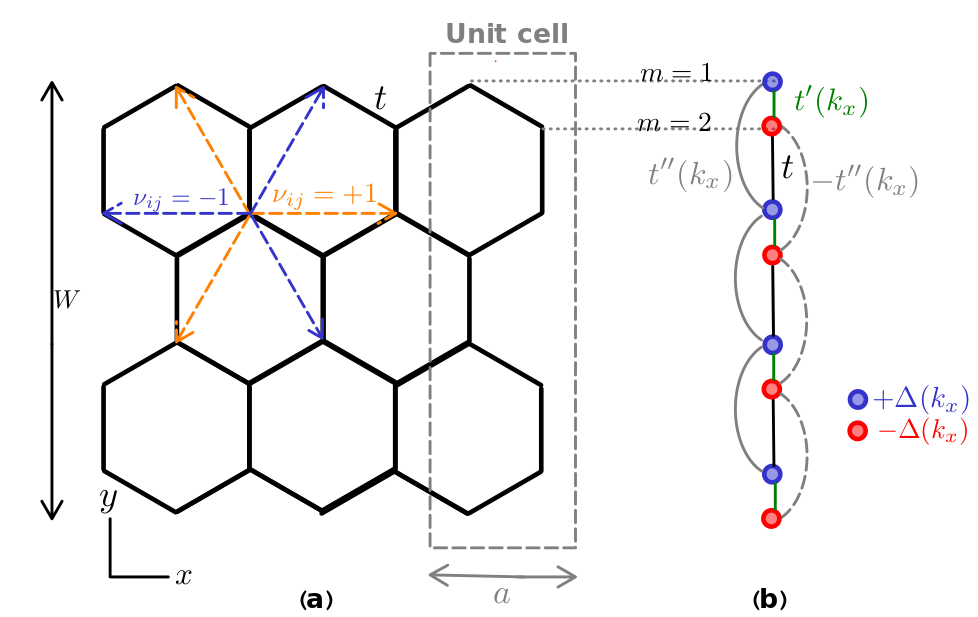}} 
        \caption{ (a) Schematic representation of a two-dimensional ribbon with honeycomb lattice structure and zigzag edges. The ribbon is infinite along the $x$ direction and finite  along the $y$ direction. The nearest-neighbor hopping parameter is $t$ and the second-nearest hopping parameter $\lambda_{\text{SO}}$ depends on the clockwise ($\nu=-1$) or counterclockwise ($\nu=+1$) direction of the hopping. The corresponding unit cell of the ribbon is shown with a rectangle with the dashed line and width $a=1$. 
        (b) The sketch of corresponding effective chain of the ribbon geometry parameterized with $k_x$ which is obtained after performing the Fourier transformation~\eqref{FT} . The momentum-dependent hopping and on-site parameters $t^{\prime}$, $t^{\prime\prime}$, and $\pm\Delta$  shown with different types of lines and colors.}
	\label{F1}
\end{figure}

	We consider the Kane-Mele model in a ribbon geometry of a two-dimensional honeycomb lattice with zigzag edges, which is shown in Fig.\ref{F1} (a). The Hamiltonian of the Kane-Mele model is~\cite{KM1,KM2}:
	\\

\begin{eqnarray}
\label{KMMod}
\mathcal{H}_{\text{KM}} &=&t\sum_{\langle i,j\rangle,\alpha }c_{i\alpha}^{\dagger
}c_{j\alpha}+i\lambda _{\text{SO}}\sum_{\langle \langle i,j\rangle \rangle
,\alpha \beta}v_{ij}c_{i\alpha}^{\dagger }s^z_{\alpha \beta}c_{j\beta}   \\ 
&+&i\lambda _{\text{R}}\sum_{\langle i,j\rangle ,\alpha \beta}c_{i\alpha }^{\dagger}(\bm{s}%
\times \bm{d}_{ij})^z_{\alpha \beta}c_{j\beta}+\lambda_\nu\sum_{i\alpha}\xi _{i}c_{i\alpha}^{\dagger
}c_{i\alpha}.  \notag 
\end{eqnarray}%
The first term describes the nearest-neighbor (NN) hopping with amplitude $t$ in which $c_{i\alpha}^\dagger$ and $c_{i\alpha}$ represent the creation and annihilation operators of an electron with spin $\alpha$ at site $i$, and $\langle i, j \rangle$ runs over all the NN sites. The second term is the intrinsic SOC with coupling $\lambda _{\text{SO}}$ between the next-nearest-neighbor (NNN) sites showed by $\langle\langle i, j \rangle\rangle$ in the summation index where $\bm{s} = (s^x, s^y, s^z)$ are the Pauli matrices for physical spins.
The factor $v_{ij}=\frac{\bm{d}_{i}\times \bm{d}_{j}}{|\bm{d}_{i}\times \bm{d}_{j}|}=\pm 1$, with $\bm{d}_{i}$ and $\bm{d}_{j}$ as the two nearest bonds connecting NNN sites $i$ and $j$, depends on the hoping path and shown in Fig.\ref{F1} (a). 
The third term represents the Rashba SOC where $\lambda _{\text{R}}$ controls the Rashba interaction strength and $\bm{d}_{ij}$ represents a unit vector pointing from the site $j$ to site $i$. The last term describes the staggered sublattice potential of strength $\lambda_\nu$ in which $\xi_i=\pm 1$ on each sublattice.

\section{Analytic perturbative approach}\label{Sec.III}
In this section, we will present our perturbative approach which allows us to derive analytical expressions for the helical edge states of the topological Kane-Mele model.
We consider two different cases: one where the Rashba SOC is absent and the other where it is present.
In the first case due to the existence of the axial spin symmetry, one can take only a single edge mode with up or down spin orientation and consider it as a chiral edge mode.
Instead, in the second case where the axial spin symmetry is broken the helical edge eigenstates are generally linear combinations of both spin-up and spin-down eigenstates.
In what follows we will derive and discuss different aspects of the above-mentioned edge states in detail.
 
\subsection{Chiral edge state wave function}

We start our approach by first considering the Hamiltonian of Eq.~\eqref{KMMod} without the Rashba term, $\lambda_R=0$ and let the staggered sublattice potential vanishes, $\lambda_\nu=0$ for simplicity.
In this regime, the model contains two copies of the Haldane model, one associated with each spin direction, and has two chiral edge states with opposite chiralities and opposite propagation directions.
Therefore, we can consider our system as two decoupled Haldane models, and to shorten the notations can drop the spin indices in this subsection. 
Let us call the right-moving mode (the edge mode moving along the positive direction) $|\psi^+_{\text{edge}}\rangle$ and its counterpart mode moving in the opposite direction and carrying the opposite spin polarization $|\psi^-_{\text{edge}}\rangle$.

As we mentioned (and shown in Fig.~\ref{F1} (a)), we will consider a horizontal ribbon geometry of the honeycomb lattice that has finite width ($W$) in the y-direction with zigzag edges and infinite length in the x-direction.
Due to the translational invariant of the system along the $x$-direction, the Bloch wave number in this direction, $k_x$, is a good quantum number. 
Therefore, we can use the following momentum representation of the electron operator as
\be
\label{FT}
c^\dagger_j =\frac{1}{\sqrt{N_x}}\sum_{k_x}  c^\dagger_{k_x} e^{-ik_x x_j},
\ee
where $N_x$ is the number of sites in the $x$-direction.
This will map the Hamiltonian of Eq.(~\ref{KMMod}) to a linear chain of sites along the $y$-direction which is shown in Fig.~\ref{F1} (b). 
After Fourier transformation, the resulting Hamiltonian can be separated into two parts (for the reason that will be clear below) as the following:
\be
\mathcal{H}(k_x) =\mathcal{H}_0(k_x)+\mathcal{H}_1(k_x),
\label{HH}
\ee
where
\bea
\mathcal{H}_0(k_x)&=&\sum\limits_{\substack{m=2 \\ m : even}}  t c^\dagger_{k_x,m+1} c_{k_x,m} \nonumber \\ 
&+& \sum_{m=1} t^{\prime\prime}(k_x) c^\dagger_{k_x,m+2} c_{k_x,m} + h.c,
\label{H0K_x}
\eea
and
\bea
\mathcal{H}_1(k_x)&=& \sum\limits_{\substack{m=1 \\ m : odd}} t^\prime(k_x) c^\dagger_{k_x,m+1} c_{k_x,m} +h.c\nonumber \\
&+& \Delta(k_x) \sum_{m=1} (-1)^{m+1}c^\dagger_{k_x,m} c_{k_x,m}.
\label{H1K_x}
\eea
In the above equations the following parameters are defined: 
$t^{\prime\prime}(k_x) = 2 \lambda_{\text{SO}} \sin(\frac{k_x}{2})$, 
$t^\prime(k_x) = 2 t \cos(\frac{k_x}{2})$,   
and
$\Delta(k_x)=2\lambda_{\text{SO}} \sin(k_x)$. 
Here we should emphasize that we considered the unit cell length to be $a$ = 1 and used the gauge transformation~\cite{Bernevig-book} $c_{k_x,m} \rightarrow e^{i\frac{k_x}{2}} c_{k_x,m}$ only on the sites located on the right side of the unit cell (with index $m=2,3,6,7,\dots$) to get rid of the factor $e^{i\frac{k_x}{2}}$ in hopping amplitudes $t^{\prime}$ and $t^{\prime\prime}$  as well.

\begin{figure}[t!]
          \center{\includegraphics[width=1\linewidth]{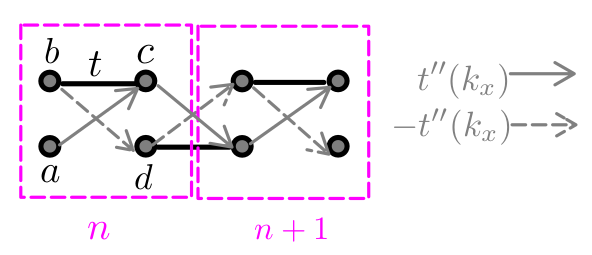}} 
        \caption{Sketch of the of two neighboring plaquettes  of the corresponding two-leg ladder system of the Hamiltonian $\mathcal{H}_0(k_x)$  obtained for the chain shown in Fig.~\ref{F1}(b). Each plaquette  consists of four different sites $a, b, c,$ and $d$.}
	\label{F3}
\end{figure}

Before proceeding further, some clarification of our approach is helpful. 
Let us take the point $k_x=\pi$ for which 
the term $\mathcal{H}_1(k_x)$ vanishes, $\mathcal{H}_1(k_x=\pi)=0$.
It is due to the vanishing of the momentum-dependent parameters $t^\prime$ and $\Delta$, namely $t^\prime(k_x=\pi) = \Delta(k_x=\pi)=0$. 
Therefore, due to the existence of the zero-energy edge states in the spectrum of the Hamiltonian $\mathcal{H}(k_x=\pi)$, the term $\mathcal{H}_0$ should provide zero-energy edge states at this point. 
Our numerical check shows that this is not only true for $k_x=\pi$, but also for any $k_x$.
It means that the edge states of the Hamiltonian $\mathcal{H}_0(k_x)$ form a zero-energy flat band.
In what follows, we first take the term $\mathcal{H}_0(k_x)$ and derive the wave function associated with its zero-energy flat band. Then, we will use this wave function to take into account the effect of $\mathcal{H}_1(k_x)$ perturbatively which makes the edge band dispersive.   

From the Hamiltonian~\eqref{H0K_x} it is obvious that 
the only relevant parameters in the Fig.~\ref{F1} are the momentum-dependent hopping amplitudes  $t^{\prime\prime}$ and $t$.  
Indeed, this system is equivalent to a two-leg ladder model shown in Fig.~\ref{F3}. 
The unit cell of this ladder is a plaquette composed of four sites $a, b, c$, and $d$.
We denote the operators $b^\dagger_n$ and $c^\dagger_n$ as the creation operators for sites on the top leg, and at the same time,  the operators $a^\dagger_n$ and $d^\dagger_n$ for the same thing on the bottom leg in the $n$th-plaquette. This allows to write the Hamiltonian of  Eq.~(\ref{H0K_x}) as the following:
\be
\mathcal{H}_0(k_x)= \sum_{n=1}^N \phi_n^\dagger H \phi_n +  \phi_{n+1}^\dagger T \phi_n ,
\ee  
where $\phi^\dagger_n=(a^\dagger_n \;\; b^\dagger_n \;\; c^\dagger_n \;\; d^\dagger_n)$ and $N$ is the total number of plaquettes.
Here, $H$ and $T$ are two $4\times4$ matrices defined as
\be
H(k_x)=\left[\begin{matrix}
0 & 0&t^{\prime\prime}(k_x) & 0\\
0 & 0& t&-t^{\prime\prime}(k_x)\\
t^{\prime\prime}(k_x) & t & 0& 0\\
0 & -t^{\prime\prime}(k_x) & 0& 0\end{matrix}\right],
\ee
and
\be
T(k_x)=\left[\begin{matrix}
0 & 0&t^{\prime\prime} (k_x)& t\\
0 & 0& 0&-t^{\prime\prime} (k_x)\\
t^{\prime\prime} (k_x)& 0 & 0& 0\\
t & -t^{\prime\prime} (k_x)& 0& 0\end{matrix}\right].
\ee

\begin{figure}[t!]
          \center{\includegraphics[width=1.1\linewidth]{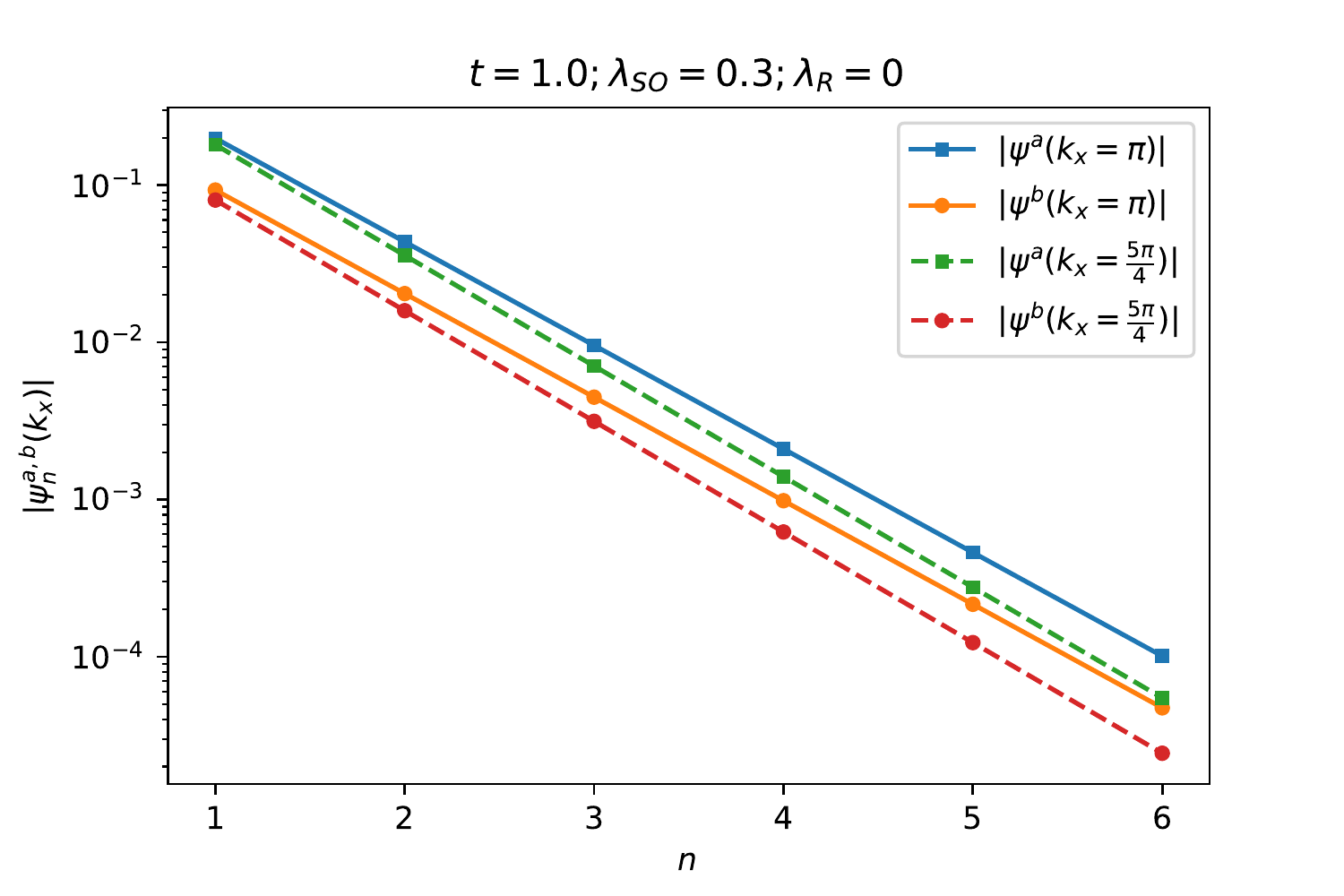}} 
        \caption{The log-linear plot of the absolute values of the non-zero amplitudes of the chiral edge states as a function of the plaquette index $n$ (shown in Fig.~\ref{F3}) for the Hamiltonian $\mathcal{H}_0(k_x)$  evaluated for two different wave numbers $k_x=\pi, 5\pi/4$. The plot with square (circle) symbols corresponds to the amplitudes of the edge states on the $a (b)$ sites.   }
	\label{FW}
\end{figure}

As stated above, we are interested in the zero-energy edge states. Thus we need to find the solution to the following Schrödinger equation 
\be
\mathcal{H}_0 (k_x)|\psi^+_\text{edge} (k_x)\rangle = 0. 
\label{Edge}
\ee
Based on our numerical check we propose a solution that has zero amplitudes on sites $c_n$ and $d_n$ which means
\be
\langle 0 | c^\dagger_n d^\dagger_n |\psi^+_\text{edge} (k_x)\rangle = 0,
\ee
where $|0\rangle$ is the fermionic vacuum state.
Therefore, the single-particle edge state of the system is given by
\be
\label{EXP}
|\psi^+_\text{edge}(k_x)\rangle  = \sum_n (\psi_n^a(k_x) a^\dagger_n+\psi_n^b(k_x) b^\dagger_n) | 0\rangle,
\ee
in which $\psi_n^{a(b)}$ is the amplitude of the edge state on the $a(b)$ site of the $n$-th plaquette.  
By inserting this wave function into the eigenvalue equation of Eq.~(\ref{Edge}) we obtain the following equation:
\be
\label{TM}
t_1 \left[\begin{matrix}
\psi_{n}^a (k_x)\\
\psi_{n}^b (k_x)\end{matrix}\right] + t_2 \left[\begin{matrix}
\psi_{n+1}^a (k_x)\\
\psi_{n+1}^b (k_x)\end{matrix}\right]=0,
\ee
where
\be
\label{t1def}
t _1 = \left[\begin{matrix}
0 & t^{\prime\prime}(k_x)\\
-t^{\prime\prime}(k_x)& t   
\end{matrix}\right],
\ee
and
\be
\label{t2def}
t_2 = \left[\begin{matrix}
t & -t^{\prime\prime} (k_x)\\
t^{\prime\prime} (k_x)& 0
\end{matrix}\right].	
\ee
Here and in the following, we drop the momentum dependence of the parameters $t_1$ and $t_2$ for notational brevity.
However, based on the Eq.~\eqref{TM}  the following recursion relation can be obtained 
\be
\left[\begin{matrix}
\psi_{n+1}^a (k_x)\\
\psi_{n+1}^b (k_x) \end{matrix}\right] = \mathcal{M} \left[\begin{matrix}
\psi_{n}^a  (k_x)\\
\psi_{n}^b  (k_x)\end{matrix}\right], 
\label{E15}
\ee
with
\be 
\label{Mdef}
\mathcal{M}=-(t^\dagger_2)^{-1} t_1,
\ee
which can be solved simply as 
\be
\left[\begin{matrix}
\psi_{n+1}^a  (k_x)\\
\psi_{n+1}^b  (k_x)\end{matrix}\right] = \mathcal{M}^n \left[\begin{matrix}
\psi_{1}^a  (k_x)\\
\psi_{1}^b  (k_x)\end{matrix}\right],
\label{REC}
 \ee
 while the initial components $(\psi_1^a,\psi_1^b)$ are known.
It is now convenient to calculate the powers of  $\mathcal{M}$ using its eigenbasis.   
In doing so we write the explicit form of the $\mathcal{M}$ using Eqs.~\eqref{t1def}, \eqref{t2def}, and \eqref{Mdef} as
\be
\mathcal{M}=
\left[\begin{matrix}
1 & -\frac{t}{t^{\prime\prime}} \\
-\frac{t}{t^{\prime\prime}}  & 1+(\frac{t}{t^{\prime\prime}})^2
\end{matrix}\right].	
\label{Mmatrix}
\ee
This matrix has two eigenvalues and two eigenvectors  respectively which can be obtained by writing $\mathcal{M}$ in terms of the Pauli matrices $\bf{\sigma}^i$  and identity matrix $I_{2\times2}$ as
 \be
 \mathcal{M} = \left[1+ \frac12 (\frac{t}{t^{\prime\prime}})^2\right]I_{2\times2}-\frac{t}{t^{\prime\prime}} \sigma^x + \frac12 (\frac{t}{t^{\prime\prime}})^2 \sigma^z.
 \ee
Therefore, the eigenvalues are the following
\be
 \Lambda_1,\Lambda_2= \left[1+ \frac12 (\frac{t}{t^{\prime\prime}})^2\right]\mp \frac{t}{t^{\prime\prime}} \sqrt{\frac14 (\frac{t}{t^{\prime\prime}})^2+1}.
 \label{EVal}
\ee
It is obvious that $\Lambda_1<1$ and $\Lambda_2> 1$ and hence  we need only to obtain the corresponding eigenvector of the $\Lambda_1$ for a convergent solution which reads as follows
 \be
 \left[\begin{matrix}
\alpha_1^a(\Theta_{k_x})\\
\alpha_1^b(\Theta_{k_x})  
\end{matrix}\right] =   \left[\begin{matrix} \sin(\frac{\Theta_{k_x}}{2})\\\\\cos(\frac{\Theta_{k_x}}{2})\end{matrix}\right],
\label{EVec}
 \ee
 where $\Theta_{k_x}=\tan^{-1}( \frac{2t^{\prime\prime}(k_x)}{t(k_x)})$.
 Now, one can immediately use $\psi_1^a(k_x)=\alpha_1^a(\Theta_{k_x})$ and $\psi_1^b(k_x)=\alpha_1^b(\Theta_{k_x})$ 
 in Eq.~\eqref{REC}
  to obtain the components of the wave function at the $n$-th plaquette as $\psi_{n+1}^a  (k_x) = (\Lambda_1(k_x))^n \sin(\frac{\Theta_{k_x}}{2})$ and $\psi_{n+1}^b  (k_x) = (\Lambda_1(k_x))^n \cos(\frac{\Theta_{k_x}}{2})$.
 Fig.~\ref{FW} shows a graphical representation of the absolute values of the wave function components $|\psi_{n+1}^a  (k_x)|$ and $|\psi_{n+1}^b  (k_x)|$ as a function of the plaquette number $n$ in the log-linear scale for two different values of the wave number $k_x=\pi$ (solid lines) and $k_x=\frac{5\pi}{4}$ (dashed lines). It is obvious that the wave function with a larger wave number decays more rapidly and hence its localization length is smaller.
 
Using the expressions obtained above for the components of the wave function, it is now possible to write the explicit expression of the edge state in Eq.~\eqref{EXP} as
\be
|\psi^+_\text{edge} (k_x)\rangle = \gamma\sum_{n=1}^N (\Lambda_1(k_x))^{n-1} [ \sin(\frac{\Theta_{k_x}}{2}) a^\dagger_n+\cos(\frac{\Theta_{k_x}}{2}) b^\dagger_n] | 0\rangle,
\ee
in which the coefficient $\gamma$ is the normalization factor of the wave function and it is easy to show that $\gamma(k_x)=\frac{1}{\sqrt{1-(\Lambda_1(k_x))^2}}$.
Before ending this subsection we should emphasize that the other edge mode $|\psi^-_\text{edge} (k_x)\rangle$ that has both the opposite spin and moving direction (left-moving mode) can be achieved by simply replacing $\lambda_{\text{SO}}$ with $-\lambda_{\text{SO}}$ in the above expressions. So far, we have obtained the edge states of the $\mathcal{H}_0(k_x)$ and will use them in the following subsection to take into account the effect of $\mathcal{H}_1(k_x)$ and derive the edge band dispersion relations.  

\subsection{Chiral edge states dispersion relation}

\begin{figure}[t!]
          \center{\includegraphics[width=1\linewidth]{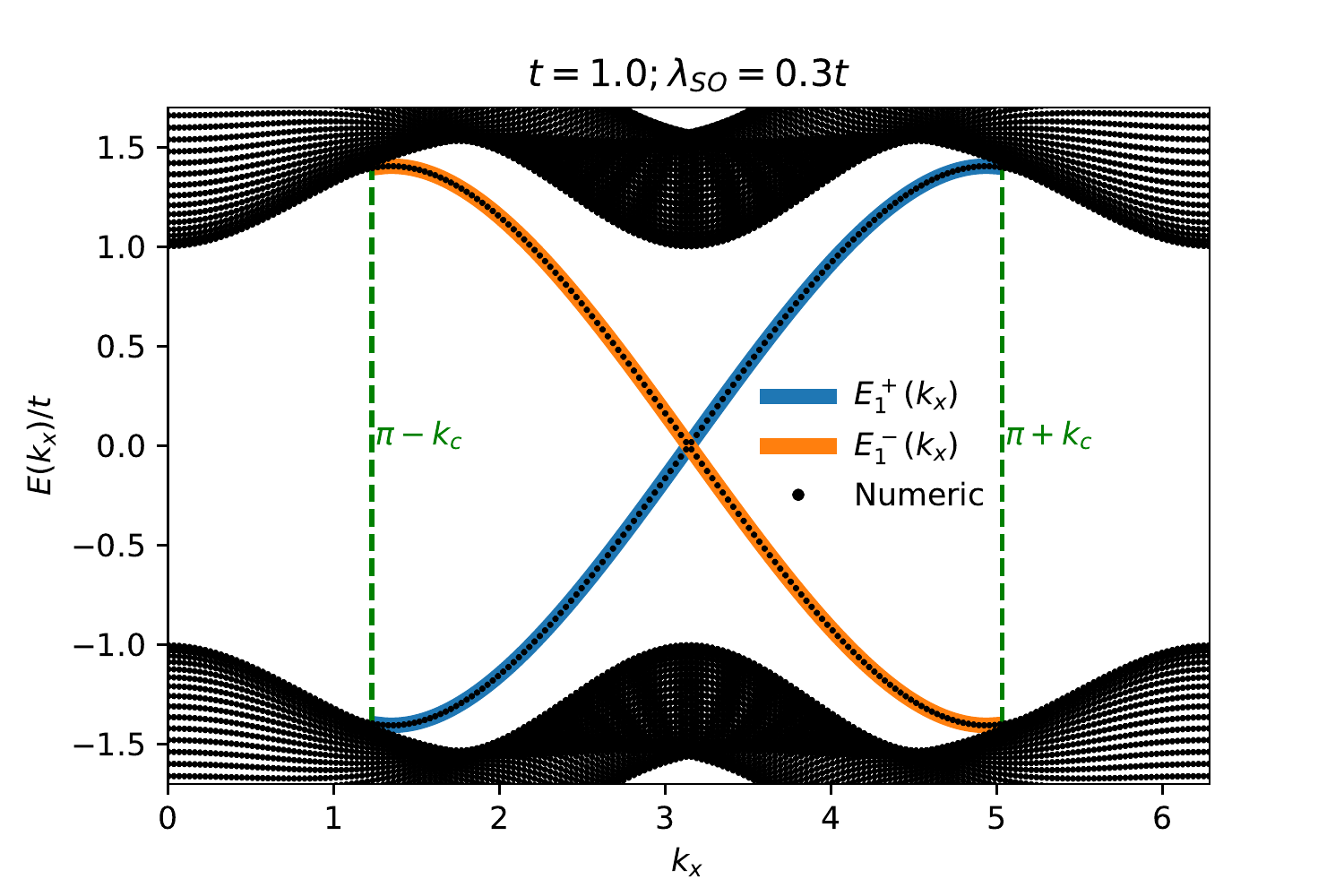}} 
        \caption{The analytical (solid lines) and numerical (points) representation of the electronic band dispersion of the chiral edge states ($E_1^+(k_x)$ and $E_1^-(k_x)$ obtained in Eq.~\eqref{Hbands}) for the Kane-Mele model with $t=1$ and $\lambda_{\text{SO}}=0.3$ in the absence of the Rashba SOC ($\lambda_R=0$). The points $k_x=\pi-k_c$ and $k_x=\pi+k_c$ denoted by the vertical dashed lines are the points where the edge bands connect the valence and conduction bands.}
	\label{FB}
\end{figure}

It is now the time to calculate the effect of considering the term $\mathcal{H}_1(k_x)$ on the flat edge band and its corresponding wave function which we obtained till now. 
Let us start to calculate the energy corrections by applying the first order perturbation theory.
The standard first-order correction in the perturbation theory can be obtained by calculating the expectation value of the perturbing 
Hamiltonian $\mathcal{H}_1(k_x)$ in the edge states of the unperturbed Hamiltonian $\mathcal{H}_0(k_x)$ as the following

\bea
E^+_1(k_x) &=& \left \langle \psi^+ _{\text{edge}} (k_x)| \mathcal{H}_1(k_x) |  \psi^+ _{\text{edge}} (k_x)\right\rangle \nonumber \\
&=&  \left[\begin{matrix}
\alpha_1^a(\Theta_{k_x}) &
\alpha_1^b(\Theta_{k_x})  
\end{matrix}\right]  
(\Delta(k_x)  \sigma^x \\ \nonumber
&&+ t^\prime(k_x)  \sigma^z)
 \left[\begin{matrix}
\alpha_1^a(\Theta_{k_x}) \\
\alpha_1^b(\Theta_{k_x})  
\end{matrix}\right].  
\eea
Using  the wave function components $\alpha_1^a$ and $\alpha_1^b$ given in Eq.~\eqref{EVec}  and after performing some algebra we will find the following 
dispersion relation for the right-moving edge state:
\be
E^+_1(k_x) = - \frac{3t\cos{(\frac{k_x}{2})}}{\sqrt{\frac{t^2}{4(t^{\prime\prime}(k_x))^2}+1}}. 
\ee
By the same token, we can obtain the energy band dispersion relation for the left-moving edge state as the following
\be
E^-_1(k_x) = \left \langle \psi^- _{\text{edge}} (k_x)| \mathcal{H}_1(k_x) |  \psi^- _{\text{edge}}(k_x) \right\rangle = -  E^+_1(k_x). 
\ee
Thus the energy dispersion relation for the chiral edge states of the Haldane model which are moving towards right and left can be written as
\be
 E^\pm_1(k_x) = \mp \frac{3t\cos{(\frac{k_x}{2})}}{\sqrt{\frac{t^2}{16 \lambda^2_{\text{SO}} \sin^2(\frac{k_x}{2})}+1}},
\label{Hbands}
\ee
correspondingly. 

Fig.~\ref{FB} shows a comparison between these analytical expressions for the band dispersions of the edge states with the numerical band structure obtained for a zigzag ribbon.
As it is obvious, our analytical expression is in excellent agreement with its corresponding numerical result for $\pi-k_c\leq k_x\leq \pi+k_c$ where $\pi\pm k_c$ are the points where the edge band touches the bulk bands.
We should note that the reason for this excellent agreement in the first-order perturbation correction is due to the large energy gap between the edge and bulk bands which results in negligible higher-order corrections. By the same reasoning, we do not need to worry about the higher order corrections to the edge states wave functions.

\subsection{Generic helical edge state analysis}
Let us now consider the effect of the Rashba spin-orbit coupling term in the Hamiltonian~\eqref{KMMod}.
It breaks the axial spin symmetry of the system and hence, the $z$-component of the spin of an electron is no longer conserved generally.
This means that for each value of the energy (or equivalently momentum) the edge states in the presence of Rashba interaction can be written as linear combinations of the spin-up state ($|\psi^+_\text{edge}(k_x)\rangle$) and spin-down state $|\psi^-_\text{edge}(k_x)\rangle$.   
Such resulting $k$-dependent edge states are called generic edge states. In what follows we will try to obtain and discuss these edge states analytically.
 
 \begin{figure}[t!]
          \center{\includegraphics[width=0.8\linewidth]{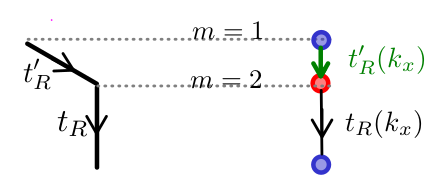}} 
        \caption{ (Left side) The illustration of the hopping terms associated with the Rashba SOC and (Right side) its corresponding momentum-dependent hopping amplitudes for the chain is shown in Fig~\ref{F1} (b).} 
	\label{FRASHBA}
\end{figure}
 
To investigate the helical edge states, we need to use the explicit spin dependence of the Rashba term in the Hamiltonian~\eqref{KMMod} which results in the new hopping parameters $t_R=i\lambda_R S^x$ and $t^\prime_R=-i\lambda_R(\sqrt{3} S^y + S^x)$ as shown in Fig.~\ref{FRASHBA}.
After performing the same Fourier transformation as in Eq.~\eqref{FT} one can immediately write the following Hamiltonian for the Rashba term:
\bea
\mathcal{H}_R(k_x)&=&\sum\limits_{\substack{m=2 \\ m : even}}  t_R(k_x) c^\dagger_{k_x,m+1} c_{k_x,m} \nonumber \\ 
&+& \sum\limits_{\substack{m=1 \\ m : odd}} t^\prime_R(k_x) c^\dagger_{k_x,m+1} c_{k_x,m} + h.c,
\label{HR}
\eea
in which $t^\prime_R(k_x)=-i\lambda_R(\sqrt{3}\sin(\frac{k_x}{2}) S^y + \cos(\frac{k_x}{2}) S^x)$ and $t_R(k_x)=t_R=i\lambda_R S^x$.
Furthermore, we need to rewrite the parameters $t, t^{\prime}, t^{\prime\prime}$ and $\Delta$ used in Eqs.~\eqref{H0K_x} and~\eqref{H1K_x} as the following
$t=tI_{2\times2}$,
$t^\prime(k_x) = 2 t \cos(\frac{k_x}{2}) I_{2\times2}$,  
$t^{\prime\prime}(k_x) = 2 \lambda_{\text{SO}} \sin(\frac{k_x}{2}) S^z$, 
and
$\Delta(k_x)=2\lambda_{\text{SO}} \sin(k_x) S^z$. 

In order to proceed further,  we use the basis states  $|\psi^+_\text{edge}(k_x)\rangle$ and $|\psi^-_\text{edge}(k_x)\rangle$ and write the Hamiltonian $\mathcal{H}_{\text{KM}}(k_x)=\mathcal{H}(k_x)+\mathcal{H}_R(k_x)$ where $\mathcal{H}(k_x)$ and $\mathcal{H}_R(k_x)$ are defined in Eqs.~\eqref{HH}  and~\eqref{HR} respectively.   
It is obvious that the Hamiltonian $\mathcal{H}(k_x)$ only contributes to the diagonal matrix elements and  $\mathcal{H}_R(k_x)$ only contributes to the off-diagonal matrix elements. Thus, one can write it as 
\be
\mathcal{H}_{\text{KM}}(k_x)=
\left[\begin{matrix}
E^+(k_x) & E_R(k_x)\\
E_R^*(k_x)& E^-(k_x)
\end{matrix}\right].	
\label{EE}
\ee 
in which $E^\pm(k_x)$ given by Eq.~\eqref{Hbands} and
\bea
E_R(k_x) &=& \langle \psi^+_\text{edge}(k_x)| \mathcal{H}_R(k_x) |\psi^-_\text{edge}(k_x)\rangle \nonumber\\
&=& -i\lambda_R\frac{\cos(\frac{k_x}{2})}{\sqrt\frac{t^2}{16\lambda_{\text{SO}}^2\sin^2(\frac{k_x}{2})+1}}.
\eea
Now we can immediately rewrite this Hamiltonian as
\be
\mathcal{H}_{\text{KM}}(k_x)= E^+(k_x) S^z + E_R(k_x) S^y,
\label{HAL}
\ee
which results in the following energy spectrum
\bea
E_{\text{KM}}^\pm(k_x) &=& \pm \sqrt{|E_R(k_x)|^2+(E^+(k_x))^2} \nonumber \\
&=& \pm\frac{\cos(\frac{k_x}{2})\sqrt{(3t)^2+\lambda_R^2}}{\sqrt\frac{t^2}{16\lambda_{\text{SO}}^2\sin^2(\frac{k_x}{2})+1}}.
\label{EEKM}
\eea

 \begin{figure}[t!]
          \center{\includegraphics[width=1.1\linewidth]{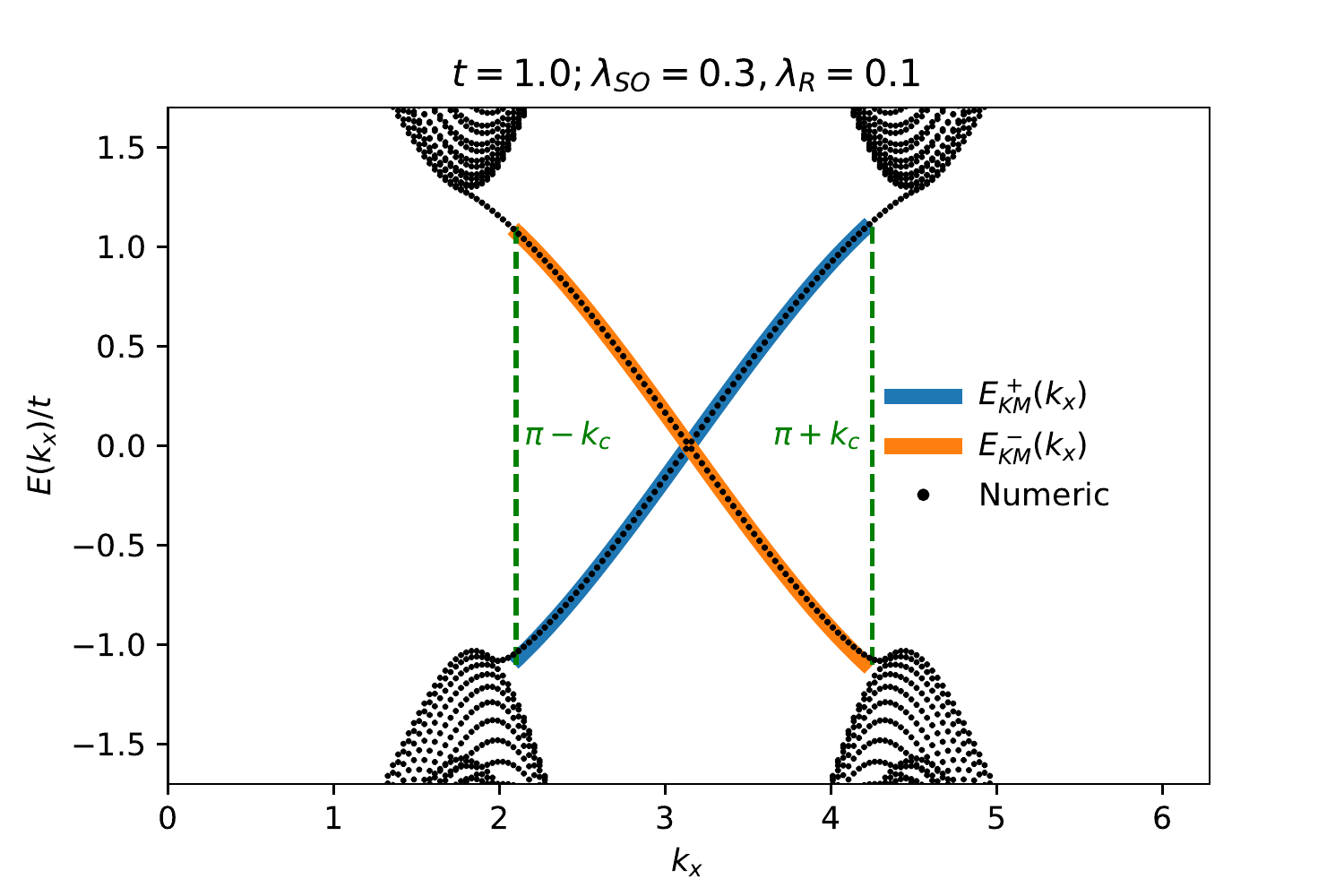}} 
        \caption{The analytical (solid lines) and numerical (points) representation of the electronic band dispersion of the helical edge states ($E_{\text{KM}}^+(k_x)$ and $E_{\text{KM}}^-(k_x)$ obtained in Eq.~\eqref{EEKM}) for the Kane-Mele model with $t=1$ and $\lambda_{\text{SO}}=0.3$ and ($\lambda_R=0.1$). The points $k_x=\pi-k_c$ and $k_x=\pi+k_c$ denoted by the vertical dashed lines are the points where the energy of the edge bands closes to the valence and conduction bands.}
	\label{BBC}
\end{figure}

Fig.~\ref{BBC} shows a graphical representation of the expressions obtained in Eq.~\eqref{EEKM} for the energy spectrum of the chiral edge modes with $t=1.0, \lambda_{\text{SO}}=0.3$ and $\lambda_R=0.1$. 
Again, it is obvious that for a wide range of wave numbers $k_x$, the analytical perturbation corrections to the edge band dispersion relation are in complete agreement with the numerical band structure obtained for a ribbon with zigzag edges. The only deviations are found near the energies close to the bulk bands. The reason for this discrepancy for very high energy states is that in the vicinity of bulk bands one needs to consider the higher-order corrections in the perturbation analysis.

\subsection{Spin rotation of the momentum eigenstates}
The last quantity of interest is the rotation of the spin of the momentum eigenstates.
This is an important issue since in the presence of the Rashba interaction the electron spin component along a fixed direction is not necessarily a good quantum number.
As we already discussed, the spins of the new edge state moving in the right and left directions can be rotated with respect to a fixed quantization axis and are generally linear combinations of the old spin-up and spin-down eigenstates.
 According to the Hamiltonian~\eqref{HAL}, it is now possible to obtain the new edge states $|\psi^\uparrow_R \rangle$ and $|\psi^\downarrow_R \rangle$ which are the edge states in the presence of Rashba term  in terms of the $ |\psi^+_\text{edge}\rangle$ and $ |\psi^-_\text{edge}\rangle$ as the following
\be
|\psi^\uparrow_R \rangle=  \cos(\frac{\Theta^\prime}{2}) |\psi^+_\text{edge}\rangle |\uparrow\rangle + i  \sin(\frac{\Theta^\prime}{2}) |\psi^-_\text{edge}\rangle |\downarrow\rangle,
\ee
and
\be
|\psi^\downarrow_R \rangle=  -\sin(\frac{\Theta^\prime}{2}) |\psi^+_\text{edge}\rangle |\uparrow\rangle + i  \cos(\frac{\Theta^\prime}{2}) |\psi^-_\text{edge}\rangle |\downarrow\rangle,
\ee
where 
\be
\tan(\Theta^\prime) = \frac{\lambda_R}{3t}.
\ee
This shows that although the spins of the right- and left-moving  states are still opposite, the rotation of the spin quantization axis of different momentum eigenstates does not depend on their momentum $k_x$. This is in contrast to the usual treatment of the generic helical edge states in which a momentum-dependent rotation of the spin of the momentum eigenstates is expected~\cite{GHS1, GHS2, GHS3, Mirlin}. However, our analysis shows that with in the first-order perturbation theory this is a constant rotation matrix.
 



\section{CONCLUDING REMARKS \label{Sec.IV}}

Despite their rather long history and important roles in the physics of topological insulators, the edge states of some well-known topological models were not derived explicitly.  
In this work, we aimed at providing a fascinating perturbative framework that allows obtaining the helical edge states of the Kane-Mele model analytically.
In our analysis, we considered two different regimes. The first regime is the one in which the Rashba SOC vanishes and hence the edge state for each spin direction can be viewed as 
a chiral edge mode. In contrast, in the second regime, the spin quantum number is no longer a good one due to the presence of Rashba SOC and the momentum eigenstates obtained as combinations of up and down spins.  
We have derived analytical expressions for the wave functions and their energy spectra of the chiral and (generic) helical edge states. 
Finally, we presented that the rotation of the spin-quantization axis takes place with a momentum-independent angle within the first order perturbation theory which we argued to be sufficient for momenta close to the time-reversal invariant point $k_x=\pi$.
Moreover, our analytical perturbative approach can in principle provide a way to obtain the closed-form expressions for the Green's function and other related transport quantities like those obtained in Refs.~\cite{Amini1, Amini2, Amini3}.

\begin{acknowledgments}
We gratefully acknowledge M. Biderang for useful discussions and comments during improvement of this work.
MA acknowledges the support of the Abdus Salam (ICTP) associateship program.

\end{acknowledgments}


\end{document}